\documentclass[aps,prb,pdflatex,10pt,showpacs,superscriptaddress]{revtex4-2}
\usepackage{amsmath}
\usepackage{amssymb}
\usepackage{amsthm}
\usepackage{bbm}
\usepackage{ulem}
\usepackage{graphicx}
\usepackage{hyperref}
\usepackage{physics}
\usepackage{systeme}
\usepackage{times}
\usepackage{epstopdf}
\usepackage{float}
\usepackage{xcolor}
\usepackage{booktabs}
\usepackage{cancel}

\hypersetup{
    colorlinks=true,linkcolor=blue,citecolor=blue,
    filecolor=blue,urlcolor=blue,breaklinks=true
}

\RequirePackage{color}

\begin{document}

\date{\today}

\title{Rotating effects on the Hall conductivity in a quantum dot}

\author{Carlos Magno O. Pereira}
\email[Carlos Magno O. Pereira - ]{carlos.mop@discente.ufma.br}
\affiliation{Departamento de F\'{\i}sica, Universidade Federal do Maranh\~{a}o, 65085-580 S\~{a}o Lu\'{\i}s, Maranh\~{a}o, Brazil}

\author{Lu\'{\i}s Fernando C. Pereira}
\email[Lu\'{\i}s Fernando C. Pereira - ]{luisfernandofisica@hotmail.com}
\affiliation{Departamento de F\'{\i}sica, Universidade Federal do Maranh\~{a}o, 65085-580 S\~{a}o Lu\'{\i}s, Maranh\~{a}o, Brazil}

\author{Denise Assafrão}
\email[Denise Assafrão - ]{denise.lima@ufes.br}
\affiliation{Departmento de Física, Universidade Federal do Espírito Santo , Vitória, ES, Brazil}

\author{Edilberto O. Silva}
\email[Edilberto O. Silva - ]{edilberto.silva@ufma.br}
\affiliation{Departamento de F\'{\i}sica, Universidade Federal do Maranh\~{a}o, 65085-580 S\~{a}o Lu\'{\i}s, Maranh\~{a}o, Brazil}

\begin{abstract}
We investigate the behavior of the quantized Hall conductivity in a two-dimensional quantum system under rotating effects, a uniform magnetic field, and an Aharonov-Bohm (AB) flux tube. By varying the angular velocity and the AB flux, we analyze their impact on the formation, shifting, and structure of quantized Hall plateaus. Our results reveal that rotation modifies the energy spectrum, leading to slight shifts in the plateau positions and variations in their widths. Additionally, we identify Aharonov-Bohm-type oscillations in $\sigma_{\text{Hall}}$, which become more pronounced for lower values of the cyclotron frequency $\omega_c$, indicating enhanced quantum interference effects in the low-field regime. These oscillations are further modulated by $\Omega$, affecting their periodicity and amplitude. The interplay between the confinement frequency $\omega_0$, the cyclotron frequency $\omega_c$, and the rotational effects plays a crucial role in determining the overall behavior of $\sigma_{\text{Hall}}$. Our findings provide insights into the interplay between rotation, magnetic field, and quantum interference effects, which are relevant for experimental investigations of quantum Hall systems in rotating systems.

\end{abstract}

\maketitle 

\section{Introduction}

 Mesoscopic systems have emerged as a prominent framework for investigating quantum phenomena that are absent in macroscopic regimes. The quantum confinement of charge carriers in low-dimensional structures gives rise to distinctive electronic \cite{nowozin2013self,PRB.1994.50.8460}, magnetic \cite{PRB.2004.69.195313,AndP.2023.535.202200371}, optical \cite{PRB.2011.84.235103,PLA.2025.534.130226}, and transport properties \cite{PRB.2006.73.045306,SR.2020.10.16280}. The interplay between geometry, external fields, and confinement potential enables the observation of persistent currents, magnetization oscillations, and quantized conductance, all of which are fundamental to the understanding of coherent quantum transport. In recent years, rotating quantum systems have attracted increasing interest as they introduce non-inertial effects into condensed matter models, offering a novel platform to probe the behavior of quantum particles in rotating frames \cite{CTP.2025.77.045106,AdP.2019.531.1900254}. These systems not only enrich the theoretical landscape but also suggest new experimental scenarios to explore the impact of rotation on phenomena such as the quantum Hall effect.

The study of rotating systems in quantum mechanics dates back to the pioneering work of Landau and Lifshitz \cite{LandauLifshitz}, who explored the effects of rotation on quantum states. In recent years, advanced experimental techniques have enabled the realization of rotating quantum systems, such as ultracold atomic gases in optical lattices and semiconductor quantum dots \cite{Dalibard2011}. These systems provide a unique platform to investigate phenomena like the Coriolis effect, centrifugal distortion, and the emergence of effective magnetic fields due to rotation. Such effects have profound implications for quantum systems' energy spectrum, wavefunctions, and transport properties, as demonstrated in studies of rotating Bose-Einstein condensates \cite{Fetter2009} and quantum Hall systems \cite{Cooper2008}.

One of the most striking phenomena in mesoscopic systems is the quantum Hall effect, where the conductivity tensor exhibits quantized plateaus under the influence of a perpendicular magnetic field. The integer quantum Hall effect, first observed by von Klitzing et al. \cite{Klitzing1980}, arises from the formation of Landau levels and the topological nature of electronic states. This discovery provided a precise method for measuring the fine-structure constant and opened new avenues for exploring topological phases of matter \cite{Thouless1982}. In recent years, investigations into rotating quantum rings have revealed additional modifications to the energy spectrum and electronic transport properties, particularly in the presence of the Aharonov-Bohm effect and spin-orbit coupling. These rotational effects have been linked to alterations in the density of states, optical absorption properties, and magnetization \cite{AP.2023.459.169547}.

The Aharonov-Bohm effect, a hallmark of quantum mechanics, has been extensively studied in mesoscopic rings and quantum dots \cite{Aharonov1959}. It manifests as a phase shift in the wavefunction of charged particles due to the presence of a magnetic flux, even in regions where the magnetic field is zero. This effect has been shown to influence persistent currents, conductance oscillations \cite{Buttiker1983}, and the energy spectrum of quantum systems \cite{FBS.2022.63.58,PE.2021.132.1144760}. When combined with rotation, the Aharonov-Bohm effect can lead to novel interference patterns and modifications in the quantum Hall conductivity, as demonstrated in recent theoretical and experimental studies \cite{PRB.1990.42.8351}.

In addition to the Aharonov-Bohm effect, the role of spin-orbit coupling in rotating quantum systems has garnered considerable interest. Spin-orbit coupling, which arises from the interaction between an electron's spin and its motion, can significantly alter the electronic properties of quantum dots and rings \cite{Winkler2003}. In the presence of rotation, spin-orbit coupling can lead to the emergence of new topological phases and edge states, as predicted in theoretical studies of rotating topological insulators \cite{Bernevig2006}. These findings highlight the rich interplay between rotation, magnetic fields, and spin-orbit coupling in mesoscopic systems.

Theoretical models of rotating quantum systems often rely on the concept of an effective magnetic field, which arises due to the Coriolis force in a rotating frame. This effective field can mimic the effects of a real magnetic field, leading to the formation of Landau levels and quantized Hall conductivity \cite{Cooper2008}. However, the combination of real and effective magnetic fields introduces additional complexity, as the resulting energy spectrum depends on the relative strength and orientation of these fields. Recent studies have explored this interplay in the context of rotating quantum dots and rings, revealing novel phenomena such as the fractional quantum Hall effect in rotating Bose-Einstein condensates \cite{Regnault2003}.

In this study, we aim to extend the analysis of rotational effects to the Hall conductivity in mesoscopic quantum rings. We adopt a solvable model based on previous works \cite{PRB.1990.42.8351}, incorporating the rotational frame effects into the Hamiltonian and deriving the corresponding energy eigenvalues. Our model considers a two-dimensional electron gas confined in a quantum dot, subjected to a uniform magnetic field, an Aharonov-Bohm flux, and rotation. By solving the Schrödinger equation in this framework, we explore how rotation modifies the quantization of the Hall conductivity and the associated transport properties. Our results provide new insights into the interplay between rotation, magnetic fields, and quantum confinement, with potential applications in quantum computing and spintronics.

The remainder of this paper is organized as follows. In Sec.~\ref{sec_model}, we present the theoretical model describing a two-dimensional electron gas confined in a quantum dot and subjected to a magnetic field, Aharonov-Bohm flux, and rotational effects. We derive the energy eigenvalues and eigenfunctions, and use them to obtain analytical expressions for the persistent current and Hall conductivity, including thermal effects through the Fermi-Dirac distribution. In Sec.~\ref{analysis}, we perform a detailed numerical analysis of the Hall conductivity, examining the influence of rotation on the formation, stability, and modulation of quantized plateaus, as well as the emergence of Aharonov-Bohm-type oscillations. We also investigate how finite temperature modifies the conductivity profile and enhances the impact of rotational effects. Finally, in Sec.~\ref{conclusion}, we summarize our main findings and discuss possible directions for future research.

\section{Theoretical Model \label{sec_model}}

We consider a two-dimensional system of non-interacting electrons subjected to a uniform magnetic field $\mathbf{B}$, a confining radial potential $V(r)$, and a rotating frame with angular velocity $\mathbf{\Omega}$. The system also incorporates the Aharonov-Bohm (AB) effect. Under these conditions, the Schrödinger equation describing the motion of a particle with effective mass $\mu$ and charge $e$ is given by \cite{DANTAS201511,Book_Rotating_Frames}
\begin{equation}
\left\{\frac{1}{2\mu}\left[\mathbf{p}-e\left(\mathbf{A}_{1}+\mathbf{A}_{2}\right)-\mu\mathbf{A}_{3}\right]^{2}-\frac{1}{2}\mu\,\Omega^{2}r^{2}+ V_{\text{ring}}\left(r\right)\right\} \psi=E\psi,\label{sh}
\end{equation}
where $\mathbf{A}_1 = (Br/2) \boldsymbol{\hat{\varphi}}$ is the vector potential associated with the uniform magnetic field, $\mathbf{A}_2 = (\ell\hbar/e r) \boldsymbol{\hat{\varphi}}$ represents the Aharonov-Bohm flux, and $\mathbf{A}_3 = \Omega r \boldsymbol{\hat{\varphi}}$ is the effective vector potential due to the rotating frame, $\ell = \varPhi/\varPhi_0$ is the AB flux parameter, with $\varPhi_0 = h/e$ being the magnetic flux quantum.
The confining potential $V_{\text{ring}}(r)$ models a two-dimensional quantum ring and is given by \cite{SST.1996.11.1635}
\begin{equation}
V_{\text{ring}}(r)=\frac{\mu \omega _{0}^{2}r_{0}^{4}}{8r^{2}}+\frac{\mu \omega _{0}^{2}}{8} r^{2}-\frac{1}{4}\mu \omega _{0}^{2}r_{0}^{2},
\label{PotRad}
\end{equation}
where $r_0$ is the average radius of the ring, and $\omega_0$ defines the strength of the transverse confinement. The first term represents a repulsive potential that prevents the electrons from collapsing to the origin. In contrast, the second term corresponds to parabolic confinement that restricts the wavefunctions to a finite region. The third term ensures that the minimum of the potential occurs at $r = r_0$. This potential describes a quantum ring and can also model other systems, such as a quantum dot when $r_0 = 0$.
The energy eigenvalues and eigenfunctions of the Schrödinger equation (\ref{sh}) are given by \cite{CTP.2024.76.105701}
\begin{equation}
E_{n,m}=\left(n+\frac{1}{2}+\frac{L}{2}\right) \hbar\omega_{1} -\frac{m-\ell}{2}\hbar \omega_{2}-\frac{1}{4}\mu \omega _{0}^{2}r_{0}^{2}, 
\label{Eq:Enm}
\end{equation}
and
\begin{equation}
\psi_{n,m} (r,\varphi ) =\frac{1}{\lambda }\sqrt{\frac{\Gamma\left[n+L+1\right]}{2\pi n!\left(\Gamma\left[L+1\right]\right)^{2}}}
\mathrm{e}^{\mathrm{i}m\varphi }\mathrm{e}^{ -\frac{r^{2}}{4\lambda ^{2}}}\left( \frac{r^{2}}{2\lambda ^{2}}\right) ^{\frac{L}{2}}
{\mathrm{M}\left( -n,\,1+L ,\frac{r^{2}}{2\lambda ^{2}}\right)},
\label{Eq:funcaodeonda}
\end{equation}
where $n = 0, 1, 2, \ldots$ is the radial quantum number, $m = 0, \pm 1, \pm 2, \ldots$ is the magnetic quantum number, $\lambda = \sqrt{\hbar / \mu \omega_1}$ is the effective magnetic length, and $L = \sqrt{(m - \ell)^2 + \mu^2 \omega_0^2 r_0^4 / 4\hbar^2}$ is the effective angular momentum. Here, $\omega_1 = \sqrt{\omega_c^2 + \omega_0^2 + 4\Omega\omega_c}$, $\omega_2 = \omega_c + 2\Omega$, and $\omega_c = eB/\mu$ is the cyclotron frequency. The function $\mathrm{M}(a, c, x)$ is the confluent hypergeometric function of the first kind.

The persistent current $I_{n,m}$ is computed using the Byers-Yang relation \cite{PRL.7.46.1961}
\begin{equation}
I_{n,m}=-\frac{1}{\phi _{0}}\frac{\partial E_{n,m}}{\partial \ell}.
\label{Byers-Yang}
\end{equation}
Substituting Eq. (\ref{Eq:Enm}) into the above expression, we obtain:
\begin{equation}
I_{n,m}=\frac{e\omega _{1}}{4\pi }\left( \frac{m-\ell }{L}-\frac{\omega _{2}}{\omega _{1}}\right). 
\label{Inm}
\end{equation}
For a quantum dot, special care must be taken for states with $m - \ell = 0$, as the wavefunction does not vanish at $r = 0$. The current carried by these states is given by
\begin{equation}
I_{n,m}=\lim_{r_{0}\rightarrow 0}\lim_{m\rightarrow \ell }\frac{e\omega_{1}}{4\pi }\left(\frac{m-\ell }{L}-\frac{\omega _{2}}{\omega_{1}}\right)=-\frac{e\omega_{2}}{4\pi }.
\label{Inm dot}
\end{equation}

The Hall conductivity $\sigma$ is defined as the ratio of the edge current to the applied voltage. The edge current is obtained by subtracting the contribution of the confining potential $\omega_0$ from the total current $I_{n,m}$, resulting in
\begin{equation}
I_{n,m(\text{edge})}=\pm \frac{e}{4\pi}\left(\omega_{1}-\omega_{1}^{\prime}\right), 
\label{Inm edge}
\end{equation}
where $\omega_{1}^{\prime} = \sqrt{\omega_c^2 + 4\omega_c\Omega}$. The Hall conductivity is then given by
\begin{equation}
\sigma =\frac{I_{n,m(\text{edge})}}{V}=\frac{e^{2}}{h}\frac{\omega _{1}-\omega _{1}^{\prime }}{\omega _{1}-\omega _{2}}\left( n+1\right),
\end{equation}
with $V = \Delta E / e$ being the effective voltage, where $\Delta E = E_{n+1,m} - E_{n,m} = \hbar \omega_1$ is the energy separation between adjacent Landau levels, and $n$ is the number of occupied Landau levels. In the absence of rotation, we recover the well-known result of the quantization of Hall conductivity in units of $e^2/h$. The inclusion of rotation in the model introduces additional physical phenomena, such as an effective field arising from the Coriolis force. The angular velocity $\mathbf{\Omega}$ contributes to the frequency $\omega_1$ and $\omega_2$, essential for characterizing the energy eigenvalues and Hall conductivity. The interaction between the magnetic field $\mathbf{B}$ and rotation alters the structure of Landau levels. As directly derived from energy, the persistent current $I_{n,m}$ is also influenced by rotation and edge states. In the Hamiltonian, a coupling between angular momentum and rotation \cite{PRB.1996.54.1877} can modify the current distribution and magnetization of the system. These effects are particularly significant in mesoscopic systems, where finite size and boundary conditions are crucial.   

The study of the Hall conductivity in mesoscopic systems also extends to finite temperatures, where thermal effects play a significant role. At finite temperatures, the thermal population of electronic states can lead to deviations from the idealized quantized behavior observed at very low temperatures. Understanding these temperature-dependent effects is crucial for practical applications, as real-world devices operate at non-zero temperatures.

To incorporate temperature effects into our model, we consider the thermal population of the Landau levels, which is governed by the Fermi-Dirac distribution \cite{PRB.2004.69.195313}
\begin{equation}
f(E_{n,m}) = \frac{1}{1 + e^{(E_{n,m} - \mu_{c}) / k_B T}},\label{FD}
\end{equation}
where $\mu_{c}$ is the chemical potential, $k_B$ is the Boltzmann constant, and $T$ is the temperature.

To evaluate the temperature-dependent Hall conductivity, we associate a partial contribution $\sigma_{n,m}$ to each quantum state. Following the definition of the persistent current in Eq.~(\ref{Inm}), we define $\sigma_{n,m}$ as the ratio between the current $I_{n,m}$ and the effective voltage $V$. This yields
\begin{equation}
\sigma_{n,m} = \frac{I_{n,m}}{V} = \frac{e^2}{h} \left( \frac{m - \ell}{L} - \frac{\omega_2}{\omega_1} \right).
\label{Sigma_nm}
\end{equation}
This definition ensures consistency with the zero-temperature limit, where only the lowest Landau levels contribute, and provides a physically meaningful way to account for the effect of partial occupation at finite temperatures. The total Hall conductivity at temperature $T$ then gives
\begin{equation}
\sigma(T) = \sum_{n,m} \sigma_{n,m} \, f(E_{n,m}).\label{Hall_temp}
\end{equation}
At low temperatures ($k_B T \ll \hbar \omega_c$), the quantization of the Hall conductivity remains well-defined, as only the lowest Landau levels are significantly occupied. However, thermal excitations populate higher Landau levels as the temperature increases, smearing the quantized plateaus. This effect can be quantified by analyzing the broadening of the Landau level occupations through the Fermi-Dirac distribution.

The inclusion of rotation is expected to influence the temperature dependence of the Hall conductivity by altering the spacing between Landau levels. Such modifications may affect the thermal occupation of states and, consequently, the overall behavior of the conductivity. In this work, we aim to investigate how the interplay between the external magnetic field, the rotation-induced effective field, and thermal excitations impacts the Hall conductivity. By incorporating finite-temperature effects into our model, we intend to explore whether the presence of rotation leads to observable changes in the temperature-induced broadening of the Landau levels and the smearing of the quantized Hall plateaus.
    
\section{Numerical analysis of the results \label{analysis}}

In this section, we analyze the behavior of the quantized Hall conductivity $\sigma_{\text{Hall}}$ under the influence of rotation. We provide numerical results for different values of $\Omega$, examining their impact on the formation and shift of quantized plateaus.

\begin{figure}[!h]
\centering
\includegraphics[scale=0.27]{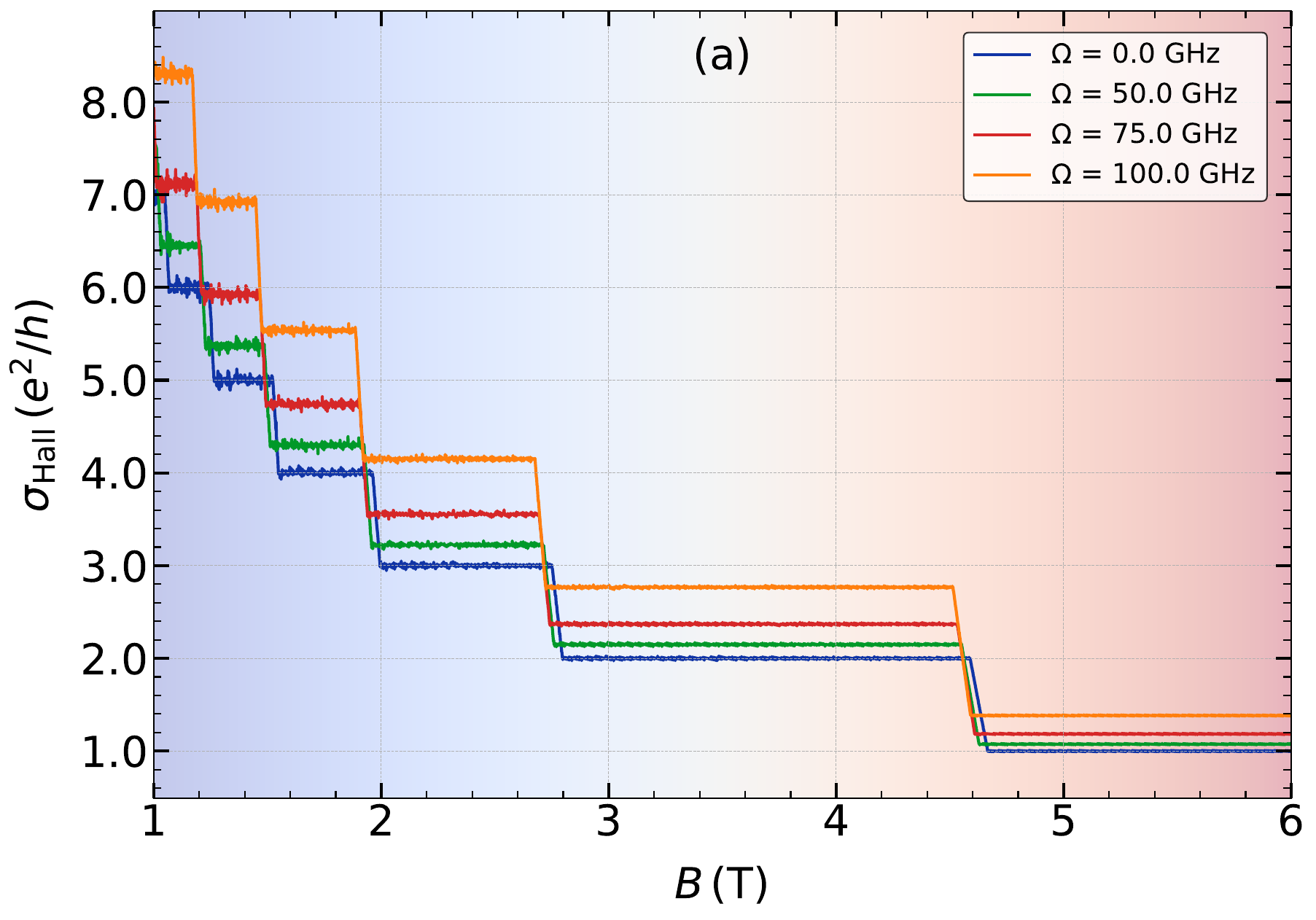}
\includegraphics[scale=0.27]{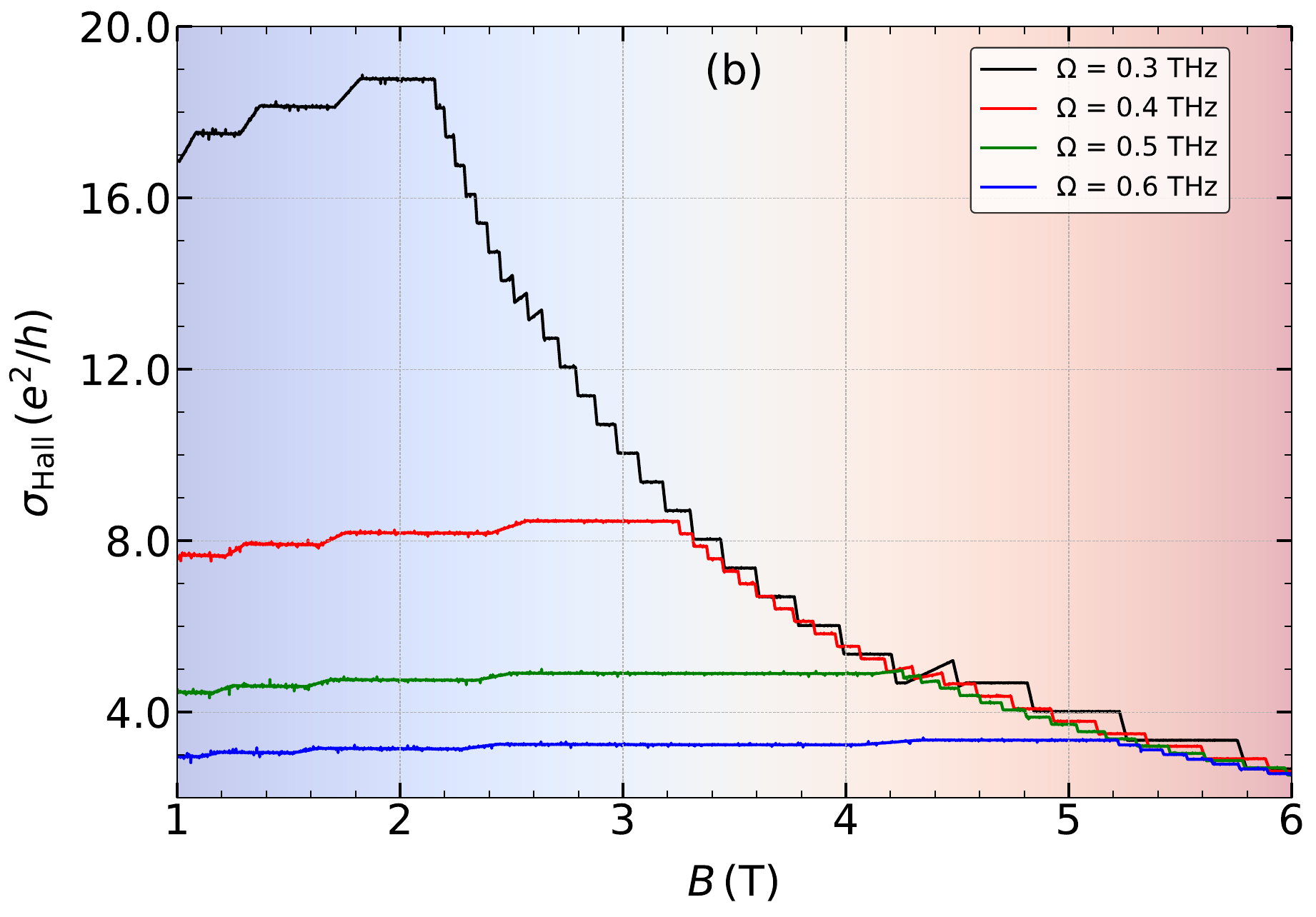}
\caption{Quantization of the Hall conductance as a function of $B~(\mathrm{T})$. Aharonov-Bohm oscillations are superimposed on the Hall plateaus. In Fig. (a), plateaus are shown for different values of $\Omega$ on the order of GHz, while in Fig. (b), they are displayed for different values of $\Omega$ on the order of THz.
}
\label{fig:hall}
\end{figure}

Our analysis considers a GaAs two-dimensional structure with effective mass $\mu = 0.067\, m_e$, where $m_e$ is the electron mass. The confinement frequency is defined as $\omega_0 = 0.20$ meV/$\hbar$, the Fermi energy is $E_F = 12.0$ meV, and $\Delta E_{n,m} = 0.7$ meV. We focus on the case of $m>0$ and examine the Hall conductivity profile as a function of $B$ under two different scenarios. In the first scenario, we investigate different values of angular velocity, specifically $\Omega = 0.0$, $50.0$, $75.0$, and $100.0$ GHz (Fig. \ref{fig:hall}-(a)). In the second case, we analyze the Hall conductivity for $\Omega=0.3$, $0.4$, $0.5$, and $0.6$ THz (Fig. \ref{fig:hall}-(b)). A key observation is that as $\Omega$ increases to a few tens of GHz, the stability of the plateaus is preserved while the magnitude of $\sigma_{\text{Hall}}$ increases. Our results further highlight the presence of AB-type oscillations along the plateaus. This effect becomes particularly significant when the magnetic field approaches $1.0$ Tesla, a regime in which quantum states become more delocalized, leading to enhanced fluctuations in $\sigma_{\text{Hall}}$. The angular velocity $\Omega$ modulates these oscillations, modifying their periodicity and amplitude. This behavior indicates an intricate interplay among $\omega_c$, $\omega_0$, and $\Omega$, which governs the Hall conductivity profile.

\begin{figure}[!t]
\centering
\includegraphics[scale=0.28]{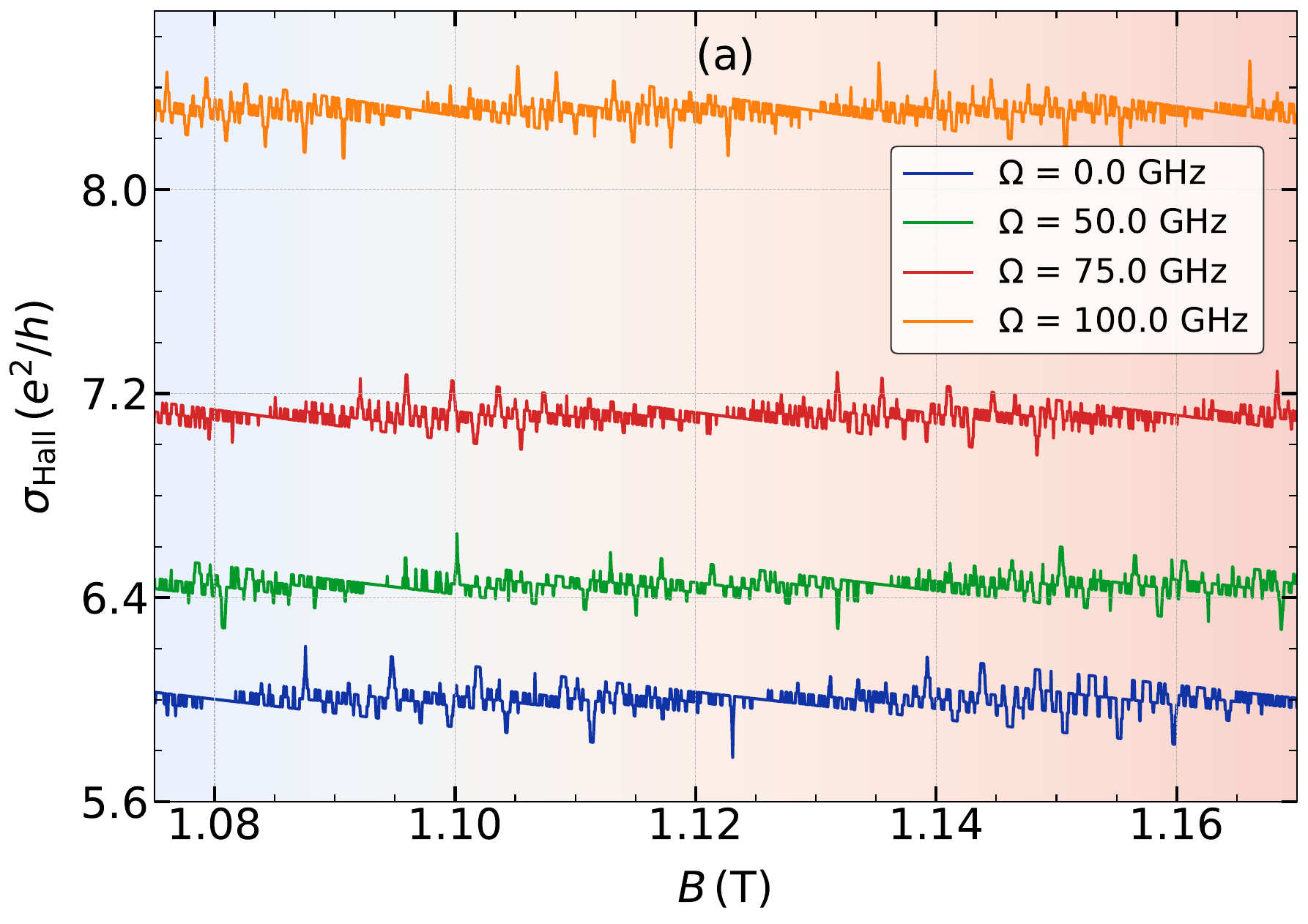}
\includegraphics[scale=0.28]{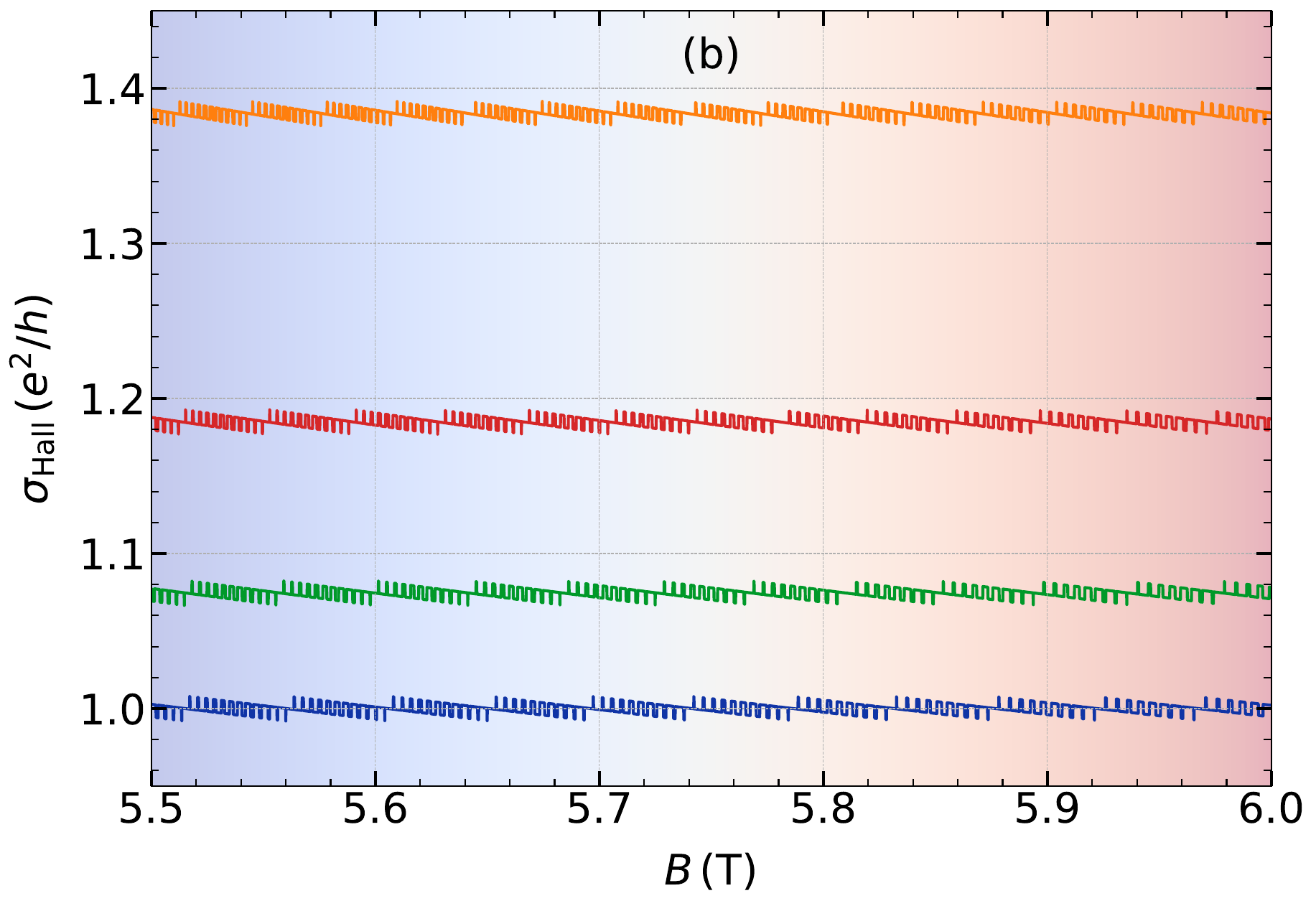}
\caption{Hall conductivity as a function of $B(\mathrm{~T})$ for different values of $\Omega$, on the order of GHz . In (a), the $B$-axis ranges from 1.08 to 1.16, while in (b), it ranges from 5.5 to 6.}
\label{fig:hall_ranges}
\end{figure}

On the other hand, when the magnitude of $\Omega$ increases to the order of THz, the plateau structure undergoes a significant transformation, leading the system into an extended oscillatory regime (Fig. \ref{fig:hall}-(b)). The fluctuations become more pronounced, indicating a strong dependence of quantum Hall states on the interplay between rotational motion and the applied magnetic field. This overall trend confirms that the rotational effects introduce an additional energy scale, which competes with both the confinement and cyclotron energies, thereby modifying the conductivity profile \cite{RP.2015.5.55,EPL.2015.110.27003,EPL.2016.116.31002}. Such behavior is characteristic of quantum interference effects in confined systems. It has been previously reported in studies of persistent currents and quantum rings subjected to external perturbations \cite{PRR.2022.4.043059}. Moreover,  the oscillatory behavior can be better understood by analyzing the conductivity within specific intervals of $B$, as shown in Fig. \ref{fig:hall_ranges}. In the range $B=1.0\,..\, 1.2$ T, the oscillations do not exhibit well-defined periodicity and amplitude, indicating a strong interplay among the various energy contributions (Fig. \ref{fig:hall_ranges}-(a)). This lack of structure suggests the system is more susceptible to rotation effects in this magnetic field regime, leading to an irregular oscillatory response. Conversely, within the range $B=5.5\,..\, 6.0$ T (Fig. \ref{fig:hall_ranges}(b)), the AB oscillations become well-defined, even for higher values of $\Omega$. This indicates that, at stronger magnetic fields, the stability of the quantum states is enhanced, reducing the impact of rotational effects and resulting in a more regular oscillatory pattern. These findings further reinforce the notion that the stability of the Hall conductivity and its oscillatory characteristics are strongly influenced by both the applied magnetic field and the angular velocity, with higher fields preserving the quantized nature of the system despite rotation perturbations.
\begin{figure}[!t]
\centering
\includegraphics[scale=0.28]{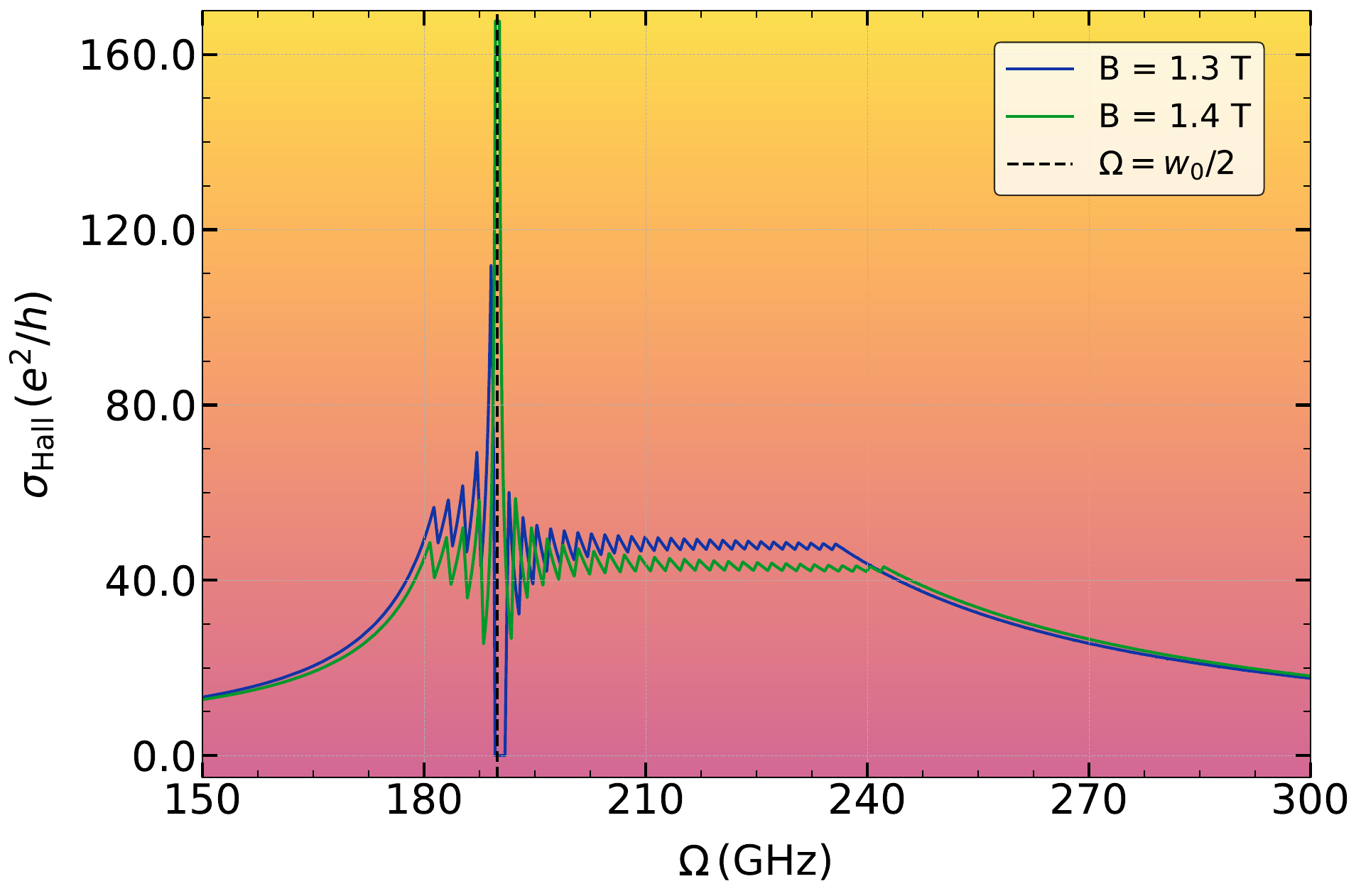}
\caption{Behavior of Hall conductance quantization as a function of \(\Omega \) (GHz) for \( B = 1.3 \) T and \( B = 1.4 \) T. The plots are divided at \( \Omega = \frac{w_0}{2} \), emphasizing the distinct behaviors on either side of this threshold.}
\label{fig3:Hall_cut}
\end{figure}
\begin{figure}[!t]
\centering
\includegraphics[scale=0.28]{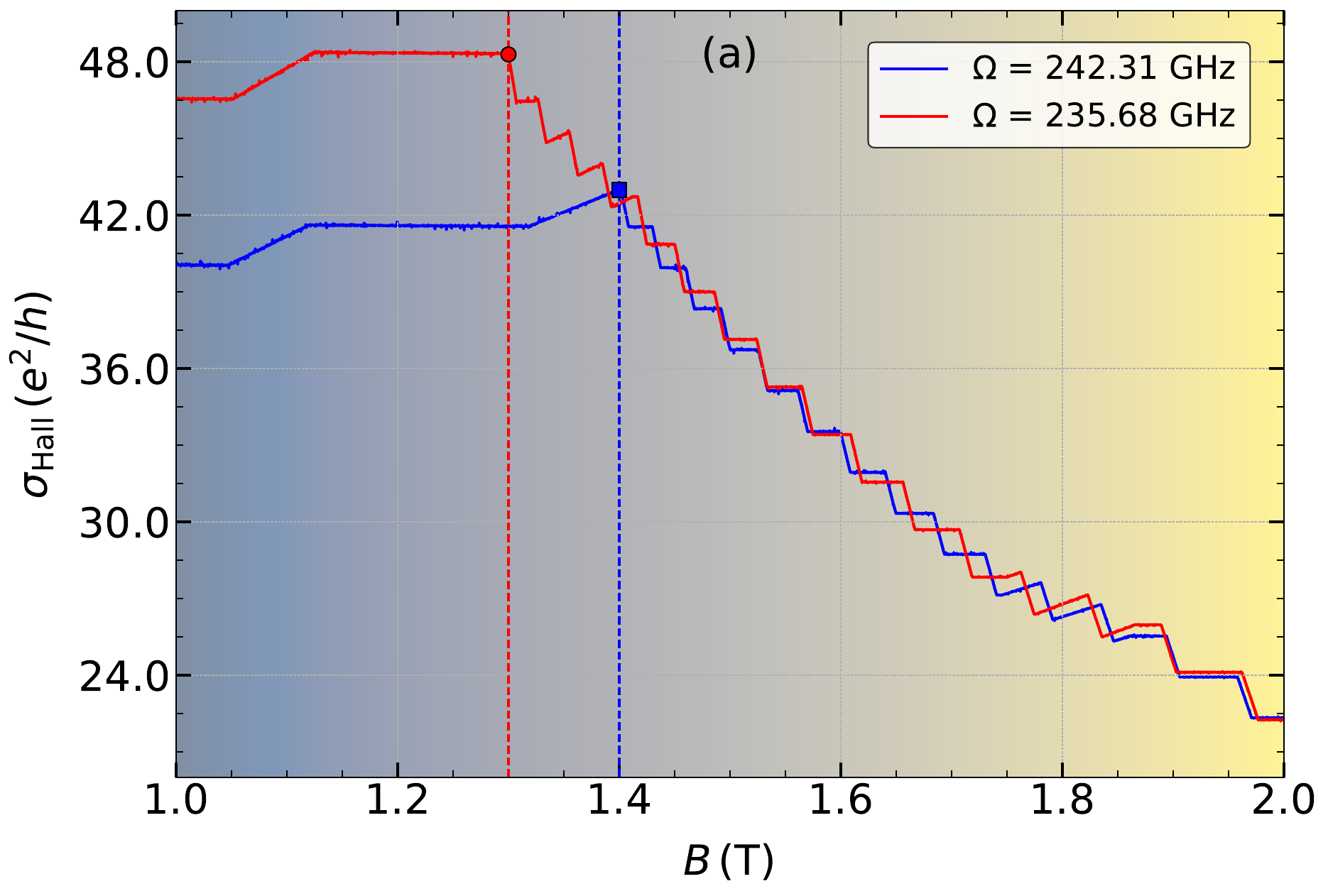}
\includegraphics[scale=0.28]{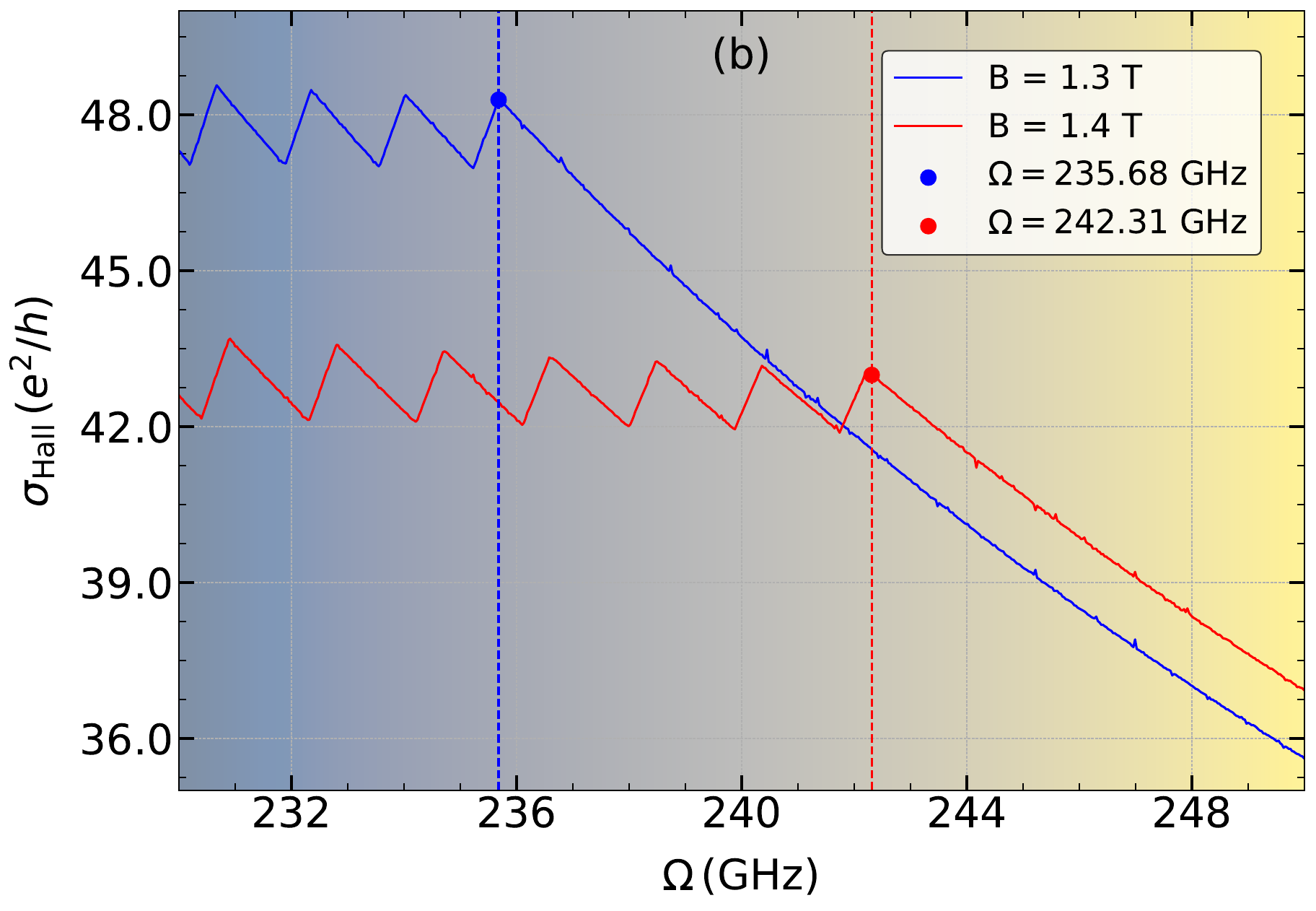}
\caption{Quantization of the Hall conductance. (a) Hall conductance as a function of \( B~(\mathrm{T}) \) for \(\Omega = 242.31\) GHz and \(\Omega = 235.68\) GHz. (b) Hall conductance as a function of \( \Omega~(\mathrm{GHz}) \) for magnetic fields \( B = 1.3 \) T and \( B = 1.4 \) T.}
\label{fig4:Hall_cut}
\end{figure}
\begin{figure}[!t]
\centering
\includegraphics[scale=0.28]{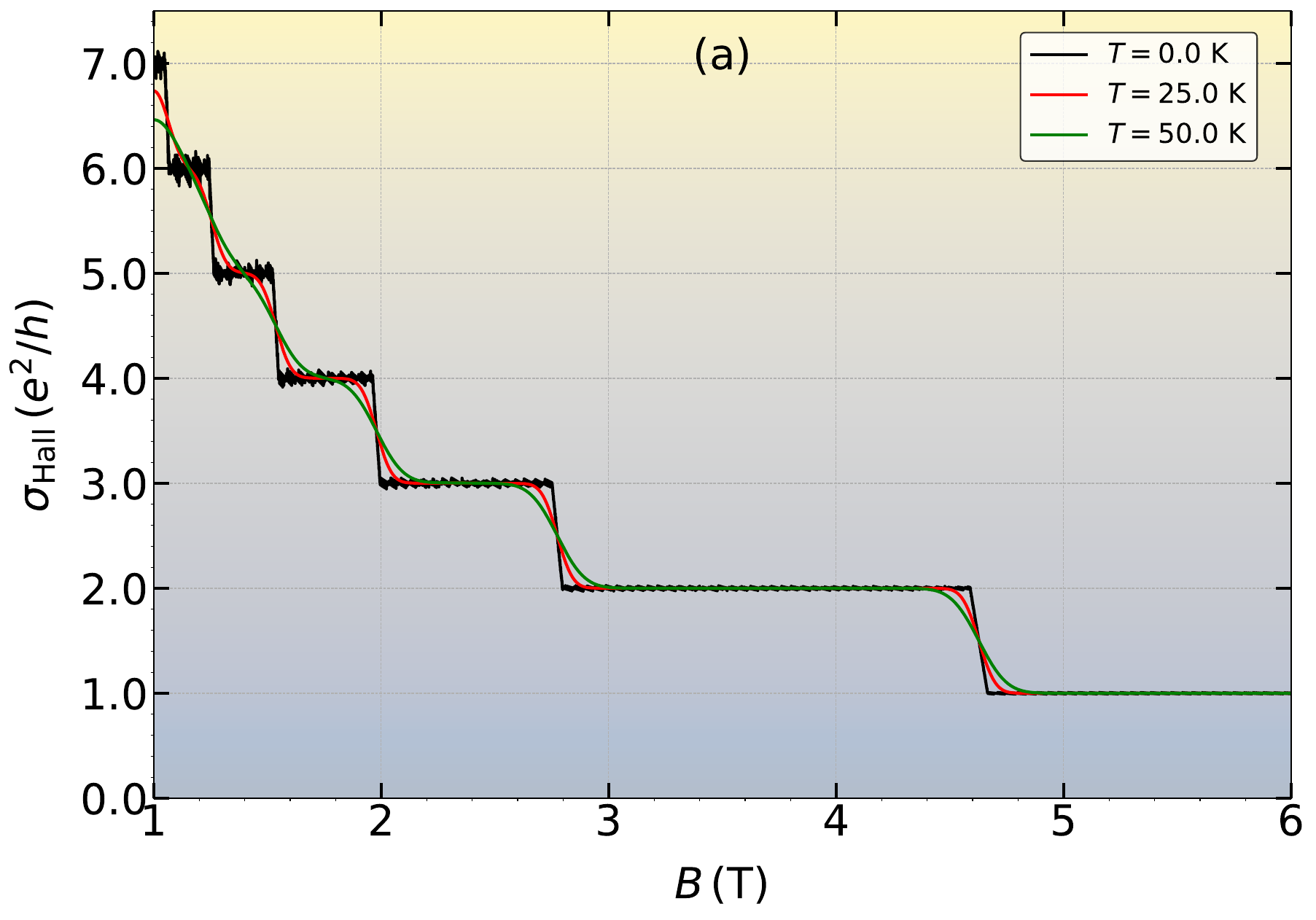}
\includegraphics[scale=0.28]{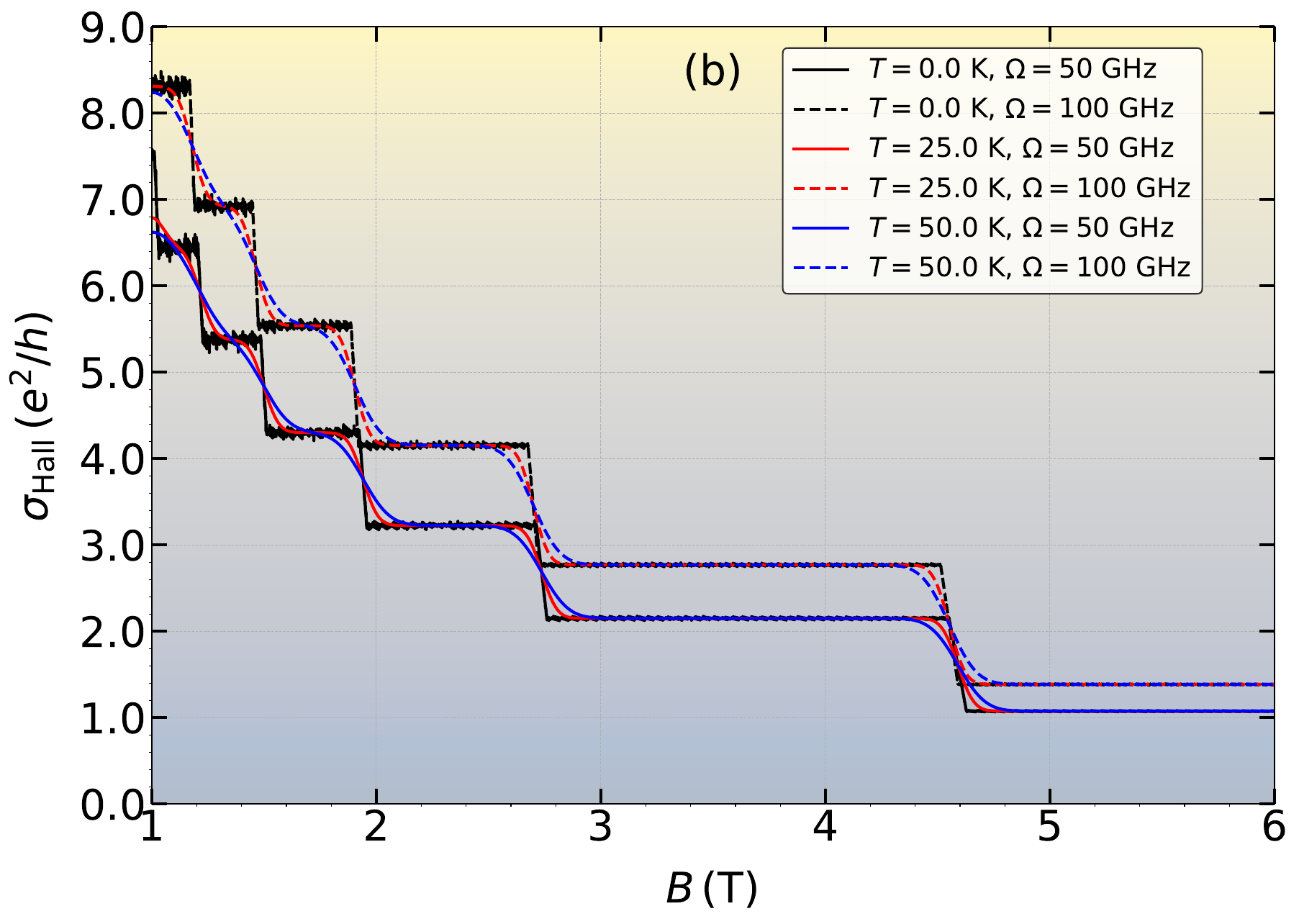}
\caption{Quantization of the Hall conductance as a function of \(B\,(\text{T})\). In (a), the results are shown for the temperatures \(T=0.0\,\text{K}\), \(T=25.0\,\text{K}\), and \(T=50.0\,\text{K}\); in (b), the same temperatures are analyzed for rotational speeds \(\Omega=50\,\text{GHz}\) and \(\Omega=100\,\text{GHz}\).}
\label{fig:Hall_Conductivity_Temperature}
\end{figure}
  
As observed in Fig. \ref{fig:hall}-(b), the profile of $\sigma_{\text{Hall}}$ as a function of $B$ changes drastically when it reaches specific values of $\Omega$. This also reveals that the behavior of $\sigma_{\text{Hall}}$ as a function of $\Omega$ exhibits significant modifications. Notably, as displayed in Fig. \ref{fig3:Hall_cut}, the results indicate that the conductivity profile undergoes substantial alterations near $\Omega = \omega_0/2$ (indicated by the black vertical dashed line), where well-defined quantum Hall plateaus break down, giving rise to emergent oscillatory behavior. This transition can be examined from two complementary perspectives: (i) its variation with respect to $B$ for fixed values of $\Omega$ (Fig. \ref{fig4:Hall_cut}-(a)), and (ii) the dependence of $\sigma_{\text{Hall}}$ on $\Omega$ for fixed values of $B$ (Fig. \ref{fig4:Hall_cut}-(b)). Figure \ref{fig4:Hall_cut}-(b) represents a specific region of Fig. \ref{fig3:Hall_cut}. In case (i), when we look at values of $\Omega$ far from $\omega_0/2$, the Hall conductivity $\sigma_{\text{Hall}}$ exhibits well-defined plateaus, characteristic of the quantum Hall effect. These plateaus correspond to stable, quantized conductivity values. However, as $\Omega$ approaches $\omega_0/2$, the plateaus progressively degrade, and an increasing number of oscillations emerge in the conductivity profile. This behavior suggests that the persistent currents, essential for sustaining quantized Hall states, become unstable in this regime. The magnetic field influences the extent of this oscillatory region: for $B = 1.3$ T (red vertical dashed line), the transition region remains relatively narrow, whereas for $B = 1.4$ T (blue vertical dashed line), the range dominated by oscillations becomes significantly broader (Fig. \ref{fig4:Hall_cut}-(a)). This trend indicates that higher values of $B$ enhance the coupling between rotational effects and the Landau level structure, thereby expanding the parameter space in which the standard quantized behavior of $\sigma_{\text{Hall}}$ is disrupted. In case (ii) (Fig. \ref{fig4:Hall_cut}-(b)), we study the $\sigma_{\text{Hall}}$ profile as a function of $\Omega$ by highlighting the rotation values ($\Omega = 235.68$ GHz (blue vertical dashed line) and $\Omega = 242.31$ GHz (red vertical dashed line)). For $\Omega$ around $\omega_{0}/2$ (Fig. \label{fig3:Hall_cut}), oscillations in conductivity appear. However, these oscillations do not have a well-defined period, and their amplitude changes with $\Omega$. We decided to classify this oscillation region as a ``Hall conductivity transition region'' coming from the rotation. A parallel interpretation of this result can be made as follows. In the $\Omega= 1.5 \,.. \,2.5$ GHz range, the quantization of $\sigma_{\text{Hall}}$ is disrupted, and pronounced oscillatory behavior emerges across a wide range of $B$. This observation confirms that as the system is driven toward $\omega_0/2$, the stability of the quantum Hall states becomes highly sensitive to fluctuations in the magnetic field. The highlighted points in the plots indicate the interval boundaries where the persistent current is not defined, indicating the transition from an ordered to a disordered conductivity regime.
Physically, this phenomenon results from the interaction between rotation and Landau quantization. The discrete Landau level structure at low $\Omega$ supports well-defined Hall plateaus. However, as $\Omega$ approaches $\omega_0/2$, alterations in the energy spectrum enable transitions between Landau levels, inducing fluctuations in $\sigma_{\text{Hall}}$. This behavior is consistent with the idea that persistent currents stabilize Hall plateaus and are disrupted near $\Omega=\omega_0/2$, giving rise to a highly oscillatory response. Furthermore, the combined effects observed in these two figures provide evidence that persistent currents become undefined near $\Omega = \omega_0/2$, leading to the breakdown of well-defined quantum Hall plateaus. Additionally, the magnetic field dependence of this instability suggests that an increase in $B$ broadens the range over which this effect occurs. These findings highlight the crucial role of rotational effects in modifying the quantum Hall regime and may have significant implications for understanding quantum transport in rotating systems.

Having analyzed the purely rotational effects on the Hall conductivity in a quantum dot, we now focus on the scenario where temperature is incorporated into the system, allowing us to explore how this additional variable combined with the rotation influences the Hall conductivity. We utilize Eq.~(\ref{Hall_temp}) to calculate $\sigma_{\text{Hall}}$ for temperatures $T = 0$, $25$, $50$~K and rotational velocities $\Omega = 50$, $100$~GHz, covering a magnetic field range of $B = 1.0$ to $6.0$~T (Fig. \ref{fig:Hall_Conductivity_Temperature}). The results exhibit a characteristic step-like quantization of the Hall conductivity, which gradually becomes smoother. This behavior is attributed to the thermal broadening of the Fermi-Dirac distribution (\ref{FD}). At $T=0$ K, the conductivity displays well-defined plateaus, characteristic of integer quantum Hall states (Fig. \ref{fig:Hall_Conductivity_Temperature}-(a)). When $T>0$, the sharpness of these plateaus diminishes, reflecting the enhanced probability of thermal excitations between Landau levels (Fig. \ref{fig:Hall_Conductivity_Temperature}-(b)). The inclusion of rotational effects modifies the energy levels through the additional terms in Eq. (\ref{Eq:Enm}), leading to shifts in the positions of the quantization steps. The rotation effects become particularly pronounced at higher temperatures, modifying the spacing between Landau levels and influencing the overall conductivity profile. Notably, for $\Omega = 100$ GHz, the Hall conductivity exhibits a more significant suppression of plateaus compared to $\Omega = 50$ GHz. This observation suggests that rotation-induced modifications to the electronic states play a crucial role in determining the transport properties of the system, as previously discussed. 

\section{Conclusion \label{conclusion}}

In this work, we investigated the Hall conductivity in rotating mesoscopic quantum dots under the influence of a magnetic field and an Aharonov-Bohm flux. The energy spectrum was obtained by solving the Schrödinger equation for a two-dimensional system with radial confinement, and the Hall conductivity was analyzed by incorporating both rotational and thermal effects.

Our results demonstrate that rotation induces an effective magnetic field, modifying the Landau level structure and significantly impacting the Hall conductivity. For moderate rotational velocities ($\Omega = 50\, ..\,100$ GHz), the quantized Hall plateaus remain stable but experience shifts in their positions and widths due to the interplay between the external magnetic field, rotation, and confinement. At higher rotational velocities ($\Omega = 0.3\, ..\,0.6$ THz), the system enters an extended oscillatory regime, where the plateaus degrade and pronounced Aharonov-Bohm-type oscillations emerge—particularly in the weak magnetic field regime ($B = 1.0\, ..\,1.2$ T), where quantum states are more delocalized and sensitive to rotational effects.

The inclusion of temperature further enriches the analysis. At zero temperature, the Hall conductivity exhibits sharp plateaus, while increasing temperature smears these features due to thermal population of Landau levels. Rotation enhances this smearing by altering the level spacing, with stronger effects observed for $\Omega = 100$ GHz compared to $\Omega = 50$ GHz.

A key finding is the breakdown of quantized Hall plateaus near $\Omega = \omega_0/2$, where persistent currents vanish, leading to strong oscillatory behavior. This transition is strongly dependent on the magnetic field, with higher values ($B = 5.5\, ..\,6.0$ T) enhancing quantum state stability and mitigating the rotational disruption.

In summary, our study shows that rotation acts as a tunable parameter in modulating the electronic and transport properties of mesoscopic systems. The emergence of oscillatory regimes and the modulation of Hall plateaus by rotation and temperature provide valuable insights into the quantum Hall effect in non-inertial frames. Future extensions could include spin-orbit coupling, electron interactions, or experimental verification, offering new perspectives for quantum devices operating in rotating or dynamic environments.

\section*{Acknowledgments}

This work was partially supported by the Brazilian agencies CAPES, CNPq, FAPES and FAPEMA. E. O. Silva acknowledges CNPq Grant 306308/2022-3, FAPEMA Grants UNIVERSAL-06395/22 and UNIVERSAL-06395/22. F. S. Azevedo acknowledges CNPq Grant No. 150289/2022-7. This study was financed in part by the Coordena\c{c}\~{a}o de Aperfei\c{c}oamento de Pessoal de N\'{\i}vel Superior - Brasil (CAPES) - Finance Code 001.

\bibliographystyle{apsrev4-2}
%

\end{document}